\documentclass[superscriptaddress, twocolumn, prl, aps]{revtex4-1}

\usepackage{graphicx}
\usepackage{amsmath, amsthm, amssymb, bm}

\begin{document}
\title{Comment on ``Edge-Induced Shear Banding in Entangled Polymeric Fluids"}

\author{Yangyang Wang}
\affiliation{Center for Nanophase Materials Sciences, Oak Ridge National Laboratory, Oak Ridge, Tennessee 37831, USA}
\author{Shi-Qing Wang}
\email{swang@uakron.edu}
\affiliation{Department of Polymer Science, University of Akron, Akron, Ohio 44325, USA}

\date{\today}
\maketitle

This comment is prompted by a recent publication by Hemingway and Fielding \cite{ref1}, who asserted that edge instability is a sufficient condition for shear banding in entangled polymeric liquids.  Unfortunately, their article did not prove that edge failure is also a necessary condition for shear banding.  Yet the authors concluded that ``true bulk banding in the absence of edge effects" is precluded, contradicting many past studies by Fielding and coworkers showing that shear banding would occur in absence of any edge effects.

The publication of the Letter \cite{ref1} is surprising because it implies that the causal relation for shear banding has been figured out for all fluids including entangled polymer liquids.   But all that this study was actually able to demonstrate was that edge effect could make the shear banding appear in the Giesekus model.  In other words, it did not show that shear banding could not occur in absence of edge effects.  Their conclusion is at odds with all past theoretical \cite{ref2, ref3}, numerical \cite{ref4}, and computer simulation \cite{ref5, ref6, ref7, ref8} studies of shear banding in polymeric and micellar solutions.

It should be noted that shear banding was experimentally observed in entangled polybutadiene solutions with cone-partitioned-plate setup where the meniscus either was eliminated by a flexible film \cite{ref9} or stayed intact \cite{ref10}. Shear banding was also prominent in soft entangled DNA solutions \cite{ref11, ref12} in absence of meniscus instabilities. Moreover, the confocal fluorescence microscopy observations of wall slip \cite{ref13} and shear banding \cite{ref14} took place more than 50 gaps away from the sample edge.

Additionally, the fracture and shear banding phenomena are sensitive to the molecular constituents of the fluids \cite{ref9, ref15}, and to the characteristics of the physical bonding \cite{ref16}.  From a molecular perspective, shear banding in polymers plausibly originates from the localized disentanglement driven by deformation \cite{ref17}.  Experimental studies have shown that the initiation of shear banding requires existence of a sizable de Gennes extrapolation length $b$, which is proportional to the ratio of the bulk viscosity to the local viscosity in the layer experiencing disentanglement, and shear banding cannot occur when $b$ is much smaller than the gap distance $H$ \cite{ref17}.  The ratio $b/H$ also controls at what Weissenberg number $Wi$ nonlinear rheological response would switch from wall slip to shear banding \cite{ref17}. Unable to describe chain disentanglement, the Letter \cite{ref1} cannot explain why insufficiently entangled polymer solutions \cite{ref18} do not show either transient or steady shear banding. Even well-entangled solutions may not display shear banding when $b$ is made sufficiently small by a proper choice of the solvent \cite{ref9}.  Admittedly, the molecular mechanism for shear banding remains elusive.

The Giesekus model \cite{ref19, ref20} was built by incorporating a stress-dependent anisotropic friction tensor into the upper-convected Maxwell model and can be derived from the kinetic theory that models polymer chains as elastic dumbbells in suspension \cite{ref21}.  This nonlinear model contains no molecular physics necessary to describe polymer entanglement.  The reason for adopting the Giesekus model in the Letter \cite{ref1} is that shear banding failed to take place in absence of edge effects according to a previous study of Fielding and coworker \cite{ref22}.

In summary, the broad claim of the Letter \cite{ref1} is not supported by its numerical simulation of the Giesekus model.  Namely, it is logically flawed to assert that bulk shear banding is precluded in the absence of edge effects without having proved that only Giesekus model is the realistic constitutive model for entangled polymeric liquids.  Edge instability is neither a sufficient nor necessary condition for shear banding.  In passing, we note that shear strain localization can take place in geometries free of meniscus, e.g., in the die entry \cite{ref23} and one-dimensional squeezing \cite{ref24} of entangled melts.  Moreover, the recent computer simulations \cite{ref6, ref7, ref8} have revealed that shear banding nucleates from molecular events of chain disentanglement without edge, in accord with the physical picture of localized yielding of  entanglement network \cite{ref17}.

\end{document}